\documentclass[aps,showpacs,prl,toolkits]{revtex4}

\usepackage{mathrsfs}
\begin{document}

\title{On the Existence of Heavy Pentaquarks: The large $N_c$ and
Heavy Quark Limits and Beyond}

\author{Thomas D. Cohen}
\email{cohen@physics.umd.edu}

\affiliation{Department of Physics, University of Maryland, College
Park, MD 20742-4111}

\author{Paul M. Hohler}
\email{pmhohler@physics.umd.edu}

\affiliation{Department of Physics, University of Maryland, College
Park, MD 20742-4111}

\author{Richard F. Lebed}
\email{Richard.Lebed@asu.edu}

\affiliation{Department of Physics and Astronomy, Arizona State
University, Tempe, AZ 85287-1504}

\begin{abstract}
We present a very general argument that the analogue of a heavy
pentaquark (a state with the quantum numbers of a baryon combined with
an additional light quark and a heavy antiquark $\overline{Q}$) must
exist as a particle stable under strong interactions in the combined
heavy quark and large $N_c$ limits of QCD\@.  Moreover, in the
combined limit these heavy pentaquark states fill multiplets of
SU(4)$\, \times \,$O(8)$\, \times \,$SU(2).  We explore the question
of whether corrections in the combined $1/N_c$ and $1/m_Q$ expansions
are sufficiently small to maintain this qualitative result.  Since no
model-independent way is known to answer this question, we use a class
of ``realistic'' hadronic models in which a pentaquark can be formed
via nucleon-heavy meson binding through a pion-exchange potential.
These models have the virtue that they necessarily yield the correct
behavior in the combined limit, and the long-distance parts of the
interactions are model independent.  If the long-distance attraction
in these models were to predict bound states in a robust way ({\it
i.e.}, largely insensitive to the details of the short-range
interaction), then one could safely conclude that heavy pentaquarks do
exist.  However, in practice the binding does depend very strongly on
the details of the short-distance physics, suggesting that the real
world is not sufficiently near the combined large $N_c$, $m_Q$ limit
to use it as a reliable guide.  Whether stable heavy
pentaquarks exist remains an open question.
\end{abstract}


\maketitle

\section{Introduction}

The existence of pentaquarks remains a vexing unresolved experimental
question.  Ten groups performing a variety of
experiments~\cite{pentaexthere,clas1,saphir} have reported the
appearance of the pentaquark state now called $\Theta^+$, a resonance
with baryon number $+1$, strangeness $+1$, and a mass in the vicinity
of 1540~MeV\@.  However, these experiments were all performed with
relatively limited statistics and significant cuts, raising the
possibility that the reported resonance is due to nothing more than
statistical fluctuations.  One ground for skepticism arises from a
series of experiments that did not find a $\Theta^+$
resonance~\cite{pentaexno,HERA-B}.  Of course, it is unclear whether
some of the experiments with negative results should be sensitive to
such an observation, since there is no reliable theoretical framework
for predicting the $\Theta^+$ production rate.  The $\Theta^+$ width
generates another source of doubt: $\Gamma(\Theta^+)$ must be
exceedingly narrow (in the range of 1--2 MeV or smaller), or it would
have been detected long ago~\cite{nussinov}, and to many it strains
credulity that such a narrow state exists in this kinematic range.

One common thread in these early reports of detection (or
non-detection) of the $\Theta^+$ is the dependence of the experimental
analysis upon revisited old data, and the appearance of the signal
only after the imposition of various cuts.  Given the limited size of
these old data sets, all of the studies yielded spectra with very
limited statistics, creating the possibility of narrow peaks due to
statistical fluctuations.  The need for high-statistics experiments
became very clear.  Special-purpose experiments designed to look for
pentaquarks with high statistics have been performed at Jefferson Lab;
the CLAS Collaboration has analyzed the high-statistics data from
photons on both a proton target~\cite{claspro} and a deuterium
target~\cite{clasde}, and finds no evidence for a $\Theta^+$ peak.
While these experiments alone do not rule out the $\Theta^+$, they
show that at least two of the previous claims of evidence for the
state, the SAPHIR $\gamma p$ result~\cite{saphir} and the CLAS $\gamma
d$ result~\cite{clas1}, were indeed statistical fluctuations.  The
prospect that other claims of evidence for the $\Theta^+$ may also
evaporate weighs heavily on the field.  The initial
observation~\cite{NA49} of $\Xi$ pentaquark states appears to be
headed for a similar fate~\cite{HERA-B}.

The theoretical landscape for pentaquarks has been just as
murky.  A paper by Diakonov, Petrov, and Polyakov~\cite{diakonov}
was seminal in focusing attention on the pentaquark, in that it
predicted a narrow state at almost exactly the mass where the
$\Theta^+$ was later reported.  However, this paper is based upon
an approximation later shown to be inconsistent with the large
$N_c$ assumptions implicit in the model~\cite{cohen}.  After the
experimental claims of pentaquarks appeared, a vast literature of
models for the $\Theta^+$ followed.  In all of these models the
existence of the $\Theta^+$ depends upon {\it ad hoc\/}
assumptions; thus they cannot be used reliably to predict the
existence of the state, and accordingly are not reviewed here.
Ultimately one may hope for lattice QCD eventually to resolve the
theoretical question of whether the state exists.  However,
current lattice simulations for both heavy and light
pentaquarks~\cite{lattice}, while always improving, still remain
inconclusive.

Given this morass, it is sensible to ask whether one can find a regime
in which the question of the pentaquark's existence is more tractable.
It has been noted previously in the context of various
models~\cite{wessling1,Lipkin} that heavy pentaquarks, states in which
the $\bar s$ quark in $\Theta^+$ is replaced by a $\bar c$ or a $\bar
b$ quark, is more likely to be bound than the $\bar s$ type.  The
principal purpose of this paper is to explore the possible existence
of heavy pentaquarks.  We show in a particular limit of QCD, the
combined large $N_c$ and heavy quark limits, that heavy pentaquarks
must exist, that they are stable under strong interactions, and that
they fall into multiplets of SU(4)$\, \times \,$O(8)$\, \times
\,$SU(2).  Here, the SU(4) is the large $N_c$ spin-flavor symmetry of
the light $u$ and $d$ quarks~\cite{gervais,others,dashen}, the O(8) is
a dynamical symmetry associated with collective vibrations of the
heavy antiquark $\overline{Q}$ (mass $m_Q$) relative to the remainder
of the system~\cite{cohensym}, and the SU(2) is the symmetry of
separate rotations of the $\overline{Q}$ spin.  We then explore the
critical question of whether $1/N_c$ and $1/m_Q$ corrections are
sufficiently small for this qualitative result to survive in the
physical world.  There are no known analytic methods starting directly
from QCD to answer this last question; thus, we investigate the
question in the context of models.

We employ models that treat the heavy pentaquark as a bound state
of a heavy meson and a nucleon interacting via pion exchange.
Although similar models have been considered
previously~\cite{shmatikov}, the present work expands on them and
is done in the context of the combined heavy quark and large
$N_c$ limits. Such models have two principal virtues: First, as
we show below, the combined large $N_c$ and $m_Q$ limit mandates
the existence of bound pentaquarks.  Indeed, our demonstration is
based on the fact that QCD in the combined limit can be reduced
to a model of this form.  Second, the long-distance behavior of
the model is well known empirically (up to experimental
uncertainties in the pion-heavy meson coupling constant).  If the
long-distance attraction due to pion exchange were sufficient to
bind the pentaquark for any reasonable choice of short-distance
dynamics (as happens in the combined limit) then one would have a
robust prediction that heavy pentaquarks exist. Unfortunately, we
find that this is not the case.

Before proceeding it is useful to clarify a semantic point.  Our
discussion relies heavily on the large $N_c$ limit of QCD; as $N_c$
becomes large, the minimum number of quarks in a baryon containing a
heavy antiquark is not 5, but rather $N_c \! + \! 2$.  Nonetheless, we
still denote such states as ``pentaquarks,'' to make the obvious
connection to the $N_c \! = \! 3$ world.

This paper is organized as follows.  In Sec.~\ref{bkgd}, we
provide a brief background on heavy pentaquarks. Section~\ref{ex}
presents a rigorous argument for the existence of heavy
pentaquarks in the combined large $N_c$ and large $m_Q$ limits.
In Sec.~\ref{sym}, we discuss the symmetry structure of heavy
pentaquarks in the combined limit, and in particular the fact
that they fall into multiplets of SU(4)$\, \times \,$O(8)$\,
\times \,$SU(2).  Then we explore in Sec.~\ref{OPEP} the question
of whether this qualitative result survives in the real world of
$N_c=3$ and finite $m_Q$ by studying simple models based on a
pion exchange between nucleons and heavy mesons.  Finally,
Sec.~\ref{concl} presents a brief discussion of the implications
of this work and concludes.

\section{Heavy Pentaquarks: Background} \label{bkgd}

The experimental situation involving reports of heavy pentaquarks
remains murky.  The H1 Collaboration at HERA has reported~\cite{h1} a
narrow resonance $\Theta_c$ appearing in $D^{*-}p$ [$(\bar c d)(uud)$]
and $D^{*+}\bar{p}$ [$(c \bar d) (\bar u \bar u \bar d)]$ states
produced in inelastic $ep$ collisions, with a mass of $3099 \! \pm \!
3 \! \pm \! 5$~MeV and a width of $12 \! \pm \! 3$~MeV\@.  We note
that the $\Theta_c$, even if it withstands further experimental
scrutiny, is {\em not\/} the type of heavy pentaquark discussed in
this paper, since it is a resonance unstable against strong decay.
Moreover, subsequent evidence argues against its existence: The FOCUS
Collaboration~\cite{focus}, using a method similar to that of H1 but
with greater statistics, finds no evidence for $\Theta_c$.  The
experimental situation for heavy pentaquarks remains in a state as
unsatisfactory as for their lighter cousins.

On the theoretical side, much of the heavy pentaquark research to
date has been performed in the context of different variants of
the quark model~\cite{wessling,wessling1,stancu}.  Our purpose
here is not to review this work in any detail, but to stress one
of its key points: Heavy pentaquarks occur far more naturally
than light pentaquarks in such models, simply because a heavy
quark is drawn more closely than a lighter quark to the bottom of
any potential well.  At the time much of the theoretical analysis
was performed, many researchers assumed that light pentaquarks
were experimentally firmly established, and so such models seemed
to make rather robust predictions of stable pentaquarks.  Now that
the existence of the light pentaquarks has become more
questionable, the reliability of heavy pentaquark predictions can
also be questioned. Nevertheless, the tendency of heavy
pentaquarks to bind more tightly than light ones remains
generically true, a simple fact that continues to play a crucial
role in the analysis of this paper.

Stewart, Wessling, and Wise~\cite{wessling1} also raise a critical
issue in the context of a diquark type model, namely, whether heavy
pentaquarks could prove stable against strong decays.  They argue that
negative-parity heavy pentaquarks should have the lowest energy (in
contrast to the positive-parity $\Theta^+$ of the Jaffe-Wilczek
model~\cite{jaffe}) since this involves s-wave interactions between
the diquarks.  They suggest that the additional attraction in such
negative-parity states might be sufficient to render the states stable
against strong decays.  In this paper we argue that pentaquarks do in
fact exist, at least in the combined large $N_c$ and large $m_Q$
limits of QCD\@.

Since the large $N_c$ limit plays a critical role in our argument, it
is useful to remark upon previous work on heavy pentaquarks as $N_c \!
\to \! \infty$.  References~\cite{wessling, JM, schat} impose large $N_c$
counting rules in the context of a quark picture as a way to implement
large $N_c$ QCD\@.  Such a picture suggests a Hamiltonian and
asymptotically stable eigenstates.  However, {\em generic\/} excited
baryons at large $N_c$ are broad resonances with $O(N_c^0)$ widths and
require an approach respecting their nature as poles occurring at
complex values in scattering amplitudes.  Two of this work's authors
have developed just such a ``scattering picture''~\cite{CL1}.  While
obtainable through a generalization of the large $N_c$ treatment for
the stable ground-state band of baryons~\cite{dashen}, the scattering
approach naturally allows a proper treatment of resonant behavior such
as large configuration mixing between resonances of identical quantum
numbers~\cite{CLconfig}.  Even for pentaquarks of $O(N_c^0)$ widths,
the scattering approach predicts multiplets degenerate in both mass
and width~\cite{CLpenta}.  But this technology, while generally true,
is not required in the current work; as we now show, the heavy
pentaquarks discussed in this paper are stable against strong decay,
at least in the combined formal limit $N_c \! \to \! \infty$, $m_Q \!
\to \! \infty$.

\section{The Existence of Heavy Pentaquarks} \label{ex}

We now show that heavy pentaquarks exist in the combined large $N_c$
and large $m_Q$ limits: They are stable against strong decay.  We must
first choose an appropriate parameter to describe the limiting
procedure.  Here, the natural choice is the $\lambda$ expansion, where
\begin{equation}
\lambda \sim 1/N_c \ , \ \Lambda_{\rm QCD}/m_Q \ ,
\end{equation}
$\Lambda_{\rm QCD}$ is the hadronic scale, and $m_Q$ is the mass of
the heavy quark.  We note that the natural expansion turns out to be
in powers of $\lambda^{1/2}$~\cite{cohensym}, instead of $\lambda^1$
for a pure 1/$N_c$ expansion.

Consider the states in the QCD Hilbert space that have energy less
than $M_N + M_H + m_{\pi}$ ($M_H$ is the mass of the lightest hadron
containing heavy antiquark $\overline{Q}$), and have baryon number
$+1$ and heavy quark $Q$ number $-1$.  These conditions exactly
describe potentially narrow heavy pentaquarks $\Theta_Q$ (assuming no
symmetry forbids the one-pion decay).  Now consider further states
with energy less than $M_N \! + \! M_H$; any pentaquark state
appearing here must be a bound state as no hadronic decay can occur.
However, scattering states clearly occur between the nucleon and the
heavy meson that have the appropriate quantum numbers and have low
enough energies.  Therefore states that can be labeled $\Theta_Q$
exist.

Yet is it possible to describe such a state as bound in some realistic
potential?  First note that momenta in the scattering states scale as
$\lambda^0$.  Therefore, since the $N,H$ reduced mass $\mu$ scales as
$\lambda^{-1}$, the kinetic energy scales as $\lambda^1$, which is
much smaller than $m_\pi \! = \! O(\lambda^0)$.  One may therefore
construct an effective theory in which all scatterings with $> \! 2$
final-state hadrons are integrated out.

However, these states naively appear nonlocal, which would prevent the
construction of a local potential.  The range of the nonlocality
scales as the inverse of momenta $p$ associated with the smallest
kinetic energy $T$ one integrates out.  In this case, $T \! \sim \!
m_{\pi}$.  Therefore, the range scales like $1/p = (2 \mu
m_{\pi})^{-1/2} \! \sim \! \lambda^{1/2} \! \to \! 0$ as $\lambda \!
\to \! 0$: The nonlocality disappears.

Next, one must ensure that the potential that binds the pentaquark
does not vanish in the combined limit.  From Witten's original $N_c$
counting~\cite{Wit}, one finds that indeed $V(\vec{r}) \sim
\lambda^0$, preventing its disappearance relative to the kinetic
energy.  Noting that the heavy quark coupling scales as $g_s \! \sim
\!  N_c^{-1/2}$, the nucleon coupling is of order $g_A/f_\pi \!
\sim \! N_c^{1/2}$, and the pion propagator is of order $m_\pi \!
\sim \! N_c^0$, one combines these ingredients to find the desired
$\lambda^0$ scaling for the potential.

We can now easily prove the existence of stable heavy pentaquarks.
Having established the locality and scaling of the potential between
heavy hadrons, we have successfully reduced a quantum field theory
problem to one of nonrelativistic quantum mechanics.  It is well
understood in this context that a potential with an attractive region
has an infinite number of bound states as $\mu \! \rightarrow \!
\infty$ (see Appendix~\ref{QM} for details).  In the present case,
$\mu \! \sim \lambda^{-1} \! \to \! \infty$, while $V(\vec{r}) \! \sim
\! \lambda^0$.  Thus, proving the existence of heavy pentaquarks in
the combined limit requires only that $V(\vec{r})$ is attractive in at
least some region.  Fortunately, we know the form of $V(\vec{r})$ at
large distances: It is given by a one-pion exchange potential (OPEP),
because $\pi$ is the lightest hadron that can be exchanged between $H$
and $N$.  It is moreover known that, regardless of the relative signs
of the coupling constants, attractive channels appear in the OPEP\@.
Thus, $V(\vec{r})$ necessarily has attractive regions, serving to bind
the heavy pentaquark.

\section{Symmetries of Heavy Pentaquarks} \label{sym}

We now show that, in the combined large $N_c$ and large $m_Q$ limit,
the pentaquark states form a multiplet of the group SU(4)$\, \times
\,$O(8)$\, \times \,$SU(2), which is an emergent symmetry of QCD\@.
The SU(4) group is a spin-flavor symmetry of the light quarks similar
to that in Refs.~\cite{gervais,others,dashen}.  The
argument~\cite{gervais,dashen} that SU(2)$_{\rm spin}
\times$SU($N_f$)$_{\rm flavor}$ combine to form a contracted
SU($2N_f$) is completely applicable to the case of heavy pentaquarks,
where here we restrict to $N_f \! = \! 2$.

The O(8) group is the symmetry associated with the configuration of
the heavy quark relative to the light degrees of freedom.  For
nonexotic baryons, the origin of this symmetry is explained in
Ref.~\cite{cohensym}.  Since the reason for such a symmetry may not be
so familiar, we provide further details here.  Consider an attractive
potential $V(\vec{r})$ of the sort described in Sec.~\ref{ex}.  Such a
$V(\vec{r})$ has a minimum, near which it can be approximated as
harmonic.  In the large $N_c$ and large $m_Q$ limits, the wave
function is localized near this minimum, creating an emergent U(3)
simple harmonic oscillator symmetry.  This U(3) symmetry is generated
by $T_{ij} \equiv a_i^\dag a_j$ $(i,j = 1,2,3)$, where $a_j$ is the
annihilation operator in the $j$th coordinate direction.  The
generators satisfy U(3) commutation relations:
\begin{equation} \label{com}
[T_{ij},T_{kl}] = \delta_{kj} T_{il} - \delta_{il} T_{kj} \ .
\end{equation}
Additionally, as $N_c \rightarrow \infty$ the creation and
annihilation operators also become generators of the emergent symmetry
with the commutation relations
\begin{equation}
[a_j,T_{kl}] = \delta_{kj} a_l \, , \quad [a_i^\dag,T_{kl}] =
-\delta_{il} a_k^\dag \, , \quad [a_i,a_j^\dag] = \delta_{ij}
\openone \, ,
\end{equation}
where $\openone$ is the identity operator.  The sixteen generators
$\{T_{ij}, \, a_l, \, a_k^\dag, \, \openone \}$ form the {\em minimal
spectrum-generating algebra\/} for the U(3) harmonic oscillator.  It
is related to the U(4) algebra generated by $T_{ij}$ $(i,j = 1,2,3,4)$
and satisfying commutation relations Eq.~(\ref{com}) by the limiting
procedure
\begin{equation}
a_j = \lim_{R \rightarrow \infty} T_{4j}/R \, , \quad a_i^\dag =
\lim_{R \rightarrow \infty} T_{i4}/R \, , \quad \openone = \lim_{R
\rightarrow \infty} T_{44}/R^2 \, .
\end{equation}
Such a procedure is called a {\em group contraction}.  Hence the group
generated by $\{T_{ij}, \, a_l, \, a_k^\dag, \, \openone \}$ is called
a contracted U(4) group.

The generating algebra of the contracted U(4) group can be expanded by
including the operators $S_{ij} = a_i a_j$ and $S_{ij}^\dag = a_i^\dag
a_j^\dag$ $(i,j = 1,2,3)$ with the following commutation relations:
\begin{eqnarray}
[S_{ij}, S_{kl}] =  [S_{ik}, a_l] & = & 0, \;\; [S_{ij},
a_k^\dag]  =  \delta_{jk} a_i + \delta_{ik} a_j , \;\; [S_{ij},
T_{kl}]  =  \delta_{jk} S_{il} + \delta_{ik} S_{jl} , \nonumber \\
{}[S_{ij}, S_{kl}^\dag] & = &
(\delta_{ki} \delta_{lj} + \delta_{il} \delta_{kj}) \openone +
\delta_{ki} T_{lj} + \delta_{lj} T_{ki} + \delta_{il} T_{kj} +
\delta_{kj} T_{li} \ ,
\end{eqnarray}
while the commutation relations for $S_{ij}^\dag$ can be obtained
through Hermitian conjugation.  This set of 28 generators
$\{S_{ij},S_{ij}^\dag,T_{ij}, \, a_l, \, a_k^\dag, \, \openone \}$
forms a closed operator algebra, which is a contracted O(8).

Reference~\cite{cohensym} continues by showing that this emergent O(8)
is also an emergent symmetry of QCD\@.  Extension to the present case
is straightforward.  The argument for the presence of the contracted
O(8) emergent symmetry relies on one's ability to approximate the
bottom of $V(\vec{r})$ as a harmonic oscillator potential.  As we have
seen, the large $N_c$ and large $m_Q$ limits ensure this feature by
leading to $\mu \! \rightarrow \! \infty$.  These conditions remain
just as true for a heavy antiquark; thus the argument
from~\cite{cohensym} applies to heavy pentaquarks.

The SU(2) is simply the symmetry of invariance under spin rotations of
the heavy quark: In the heavy quark limit, states with any alignment
of the heavy quark spin are degenerate.

\section{Bound States and the One-Pion Exchange Potential} \label{OPEP}

Now that we have shown stable heavy pentaquarks exist in the combined
large $N_c$ large $m_Q$ limit, the critical question becomes whether
they also occur in our $N_c \! = \! 3$ finite $m_Q$ world.  To our
knowledge, this question cannot be answered in a model-independent way
without solving QCD, and so we resort to models for enlightenment.

We focus here on effective potential models based upon one-pion
exchange at long distance.  As discussed in Sec.~\ref{ex}, such models
are clearly useful not only because they represent physically correct
phenomenology, but also guarantee stable pentaquarks in the combined
limit.  But we also note that the argument does not depend upon the
particular short-distance behavior of the effective potential.  If the
real world is sufficiently close to the combined-limit world for the
argument to remain valid, all models of this sort must yield
(multiple) stable pentaquarks.  Note that the masses of the various
pentaquark states can depend sensitively upon the details of the
short-distance interaction, but their existence cannot.  The question
then becomes whether models of this type predict bound pentaquarks in
a robust way, independent of the details of the short-distance
physics.  If so, one has a strong reason to believe that the states
are, in fact, bound in nature.

We construct a ``realistic'' potential that has the correct
long-distance behavior (OPEP) and an {\it ad hoc\/} short-distance
part constrained only by the natural scales of strong interaction
physics.  Our potential acts between a nucleon and a heavy meson ($D$
or $B$).  The nucleon-pion analogue is well understood; its
interaction Lagrangian reads
\begin{equation}\label{eq:pion1}
\mathcal{L}_{NN\pi} = - \frac{g_A}{f_\pi \sqrt{2}} \bar{N} \tau^a
\gamma_{\nu} \gamma_{5} N \, \partial^\nu \pi^a \ ,
\end{equation}
where the axial coupling constant $g_A \! \simeq \! 1.27$, and the
pion decay constant $f_\pi \! \simeq \!131$~MeV\@.

However, the heavy meson-pion interaction is not as straightforward.
We use a formalism similar to that outlined by Manohar and
Wise~\cite{manohar} to encode the heavy quark symmetry.  In the limit
of $N_Q$ heavy quark flavors, QCD develops an emergent SU($2N_Q$)
symmetry~\cite{isgur}.  As a consequence of this symmetry, physical
states do not depend on the spin of the heavy quark; thus the $D(B)$
and $D^*(B^*)$ mesons form a degenerate multiplet in the $m_c(m_b) \!
\to \! \infty$ limit.  The heavy meson-pion interactions
can involve transformations between pseudoscalar and vector states ($B
\leftrightarrow B^*$ or $D \leftrightarrow D^*$).  Using heavy quark
symmetry, one combines them into a single field:
\begin{equation}\label{eq:hdef}
H \equiv \frac{(1+v\!\!\!/)}{2}[P^{*}_{\mu} \gamma^{\mu} -
P\gamma_{5}] \ ,
\end{equation}
where $v^{\mu}$ is the four-velocity, and the pseudoscalar and vector
heavy meson fields are $P$ and $P^{*}_{\mu}$, respectively.  This
combination allows the interaction Lagrangian to be written in a
manner similar to that of the nucleon interaction,
\begin{equation}\label{eq:lh}
\mathcal{L}_{\rm int}=-\frac{g_{H}}{f_\pi \sqrt{2}} {\rm Tr}
\overline{H} \tau^a \gamma_{\mu} \gamma_{5} H \, \partial^{\mu}
\pi^a
\ .
\end{equation}
Of course, the pseudoscalar and vector mesons are {\em not\/}
degenerate in the real world, due to $1/m_Q$ corrections.  The mass
difference must be included in realistic models.

Both the nucleon and heavy-meson interactions with the pion can
expressed in terms of the spin and isospin of the particles:
\begin{equation}\label{eq:spinN}
\mathcal{L}_{NN\pi}=\frac{2\sqrt{2}g_A}{f_\pi} (\vec{S}_N \cdot
\vec{\partial}\pi^a) I_N^a \ ,
\end{equation}
\begin{equation}\label{eq:spinH}
\mathcal{L}_{\rm int}=\frac{2\sqrt{2}g_H}{f_\pi} (\vec{S}_l \cdot
\vec{\partial}\pi^a) I_H^a \ ,
\end{equation}
where $\vec{S}_N$ and $\vec{I}_N$ are the spin and isospin of the
nucleon, $\vec{S}_l$ is the spin of the light quark in $H$, and
$\vec{I}_H$ is the isospin of the $H$ field.  Combining
Eqs.~(\ref{eq:spinN}) and (\ref{eq:spinH}), treating the nucleon and
heavy meson in the static limit ({\it i.e.}, ignoring recoil, which is
suppressed in the combined limit) and Fourier transforming yields the
OPEP in position space:
\begin{eqnarray}\label{eq:opep}
V_\pi (\vec{r})&=& \vec{I}_N \! \cdot \vec{I}_H \, [2 S_{12} V_T (r) +
4 \vec{S}_N \cdot \vec{S}_l V_c(r)]\nonumber\\ &=&
(I^2-I_N^2-I_H^2)[S_{12} V_T (r) + (K^2-S_N^2-S_l^2) V_c (r)] \ ,
\end{eqnarray}
where the central part of the potential ($r$ measured in units of
$1/m_\pi$) is
\begin{equation}\label{eq:vc}
V_c(r)= \frac{g_A g_{H}}{2\pi f_\pi^{2}} \frac{e^{-r}}{3r} \ ,
\end{equation}
and the tensor part is
\begin{equation}\label{eq:vt}
V_T (r)=\frac{g_A g_{H}}{2\pi f_\pi^{2}} \frac{e^{-r}}{6r}
\left( \frac{3}{r^{2} } + \frac{3}{r} + 1 \right) \ .
\end{equation}
$I$ is the total isospin of the combined system, while $\vec{K} \!
\equiv \! \vec{S}_N + \vec{S}_l$, and
\begin{equation}
S_{12} \equiv 4 \, [ 3 \, (\vec{S}_N \! \cdot \hat{r})
(\vec{S}_l \cdot \hat{r}) - \vec{S}_N \! \cdot \vec{S}_l ] \ .
\end{equation}
It remains unknown whether $g_A$ and $g_H$ are of the same sign or of
different signs, so the potential could have an additional overall
negative sign.

Clearly, the OPEP dominates the interaction at large $r$ since $\pi$
is the lightest hadron.  At shorter ranges the OPEP is no longer
dominant and the effective potential is qualitatively different.  The
value of $r$ at which the OPEP ceases to dominate the effective
potential is presumably of order $1/\Lambda_{\rm QCD} \! \sim \!
1$~fm, the characteristic range in strong interactions.  Therefore,
for distances less than some cutoff value $r_0 \! \sim \! 1$~fm, we
use a purely phenomenological potential.  Note that we do not simply
add such a short-range potential to the OPEP at short distances, but
entirely replace the OPEP by this new potential: The $1/r^3$ behavior
of the tensor part of the OPEP at short ranges is unphysical and would
completely dominate the potential if not removed.  The short-distance
potentials used are taken to be either (central) constants or
quadratic functions, and their strengths are allowed to vary.  If the
logic of our underlying argument based upon the combined limit also
holds for realistic $m_Q$ values and $N_c \! = \! 3$, then the precise
details of the potentials should be irrelevant to whether the
pentaquark states bind.

We use the OPEP of Eq.~(\ref{eq:opep}) in a nonrelativistic
Schr\"odinger equation and solve for bound states.  Since the tensor
term in the potential allows mixing between $L$ states, $L$ is not a
good quantum number.  However, $S_{12}$ commutes with the parity
operator, making $P$ a good quantum number.  Therefore, states labeled
by $J$, $S$ (total spin $\vec{S} \! \equiv \! \vec{S}_Q + \!
\vec{K}$), and $P$ are used as eigenstates.  Treating states mixed
under $L$ requires a coupled-channel calculation; we obtain the
coupled equations by including all possible states labeled by $L$ and
$K$ that are consistent with a given set of $J$, $S$, and $P$.

Lastly, since this potential is intended to be ``realistic'', in
principle $B$-$B^{*}$ and $D$-$D^{*}$ mass differences can affect the
results.  Of course, these differences are $1/m_Q$ effects and vanish
in the heavy quark limit.  Since the principal reason for the model
calculation is to test qualitatively whether we live in the regime of
validity of the combined $1/N_c$ and $1/m_Q$ expansion, it makes sense
to include this difference.  However, in practice the effect of this
mass difference is entirely repulsive, making the states are less
likely to bind.  Thus, if the states do not bind in the equal-mass
case, they do not bind at all.  Accordingly, we use equal masses and
only investigate the effect of the mass splitting in cases where
binding occurs.

We attempt to make our model as realistic as possible, given the
rather simple forms assumed for the short-distance potential.  To
this end, we choose for the heavy-meson coupling constant $g_H \!
\approx \! \pm$0.59 (extracted from $D^* \! \to \! D \pi$
decay~\cite{cleo}, see below) and collect values for other
observables~\cite{pdg} in Appendix~\ref{tables},
Table~\ref{tab:const}.
As an initial guess, we also constrain the parameters of the
short-range potential such that this potential combined with a
OPEP between nucleons gives the correct 2.2~MeV deuteron binding
energy. This choice is not necessary, but it has the virtue of
ensuring that the potential parameters are not completely
unreasonable from the point of view of hadronic physics.  We
summarize the potentials in Table~\ref{tab:potential}.
Ultimately, we vary many of the parameters in order to probe the
robustness of the qualitative results.

We then solve coupled differential equations using standard
numerical methods.  We seek bound-state solutions for all $J \! =
\! \frac{1}{2}$ and $J \! = \! \frac{3}{2}$ states using both a
constant and a quadratic form for the short-distance potential, for $I
\! = \! 0$ and $I \! = \! 1$, and with either sign of $g_H$ relative
to $g_A$.  Initially (as discussed above), we assume no mass splitting
between the pseudoscalar and vector mesons.  A complete set of tables
of bound states thus obtained appears in Tables~\ref{tab:Bmeson} and
\ref{tab:Dmeson}. Here we focus on describing some key features
of these results.

For constant and quadratic potentials constrained by matching to
the deuteron energy, bound states of the pentaquark are quite
sparse. No channel supports a bound state with a $D$ meson.  The
$B$ meson is able to bind weakly in the channels with negative
parity, but only with $I \! = \! 0$.  Binding in these states is
relatively weak, around 1.3~MeV for the constant potential and
around 3.9~MeV for the quadratic potential, and binding energies
are consistently the same between these channels
(Table~\ref{tab:Bmeson}, Cols.\ A and B). It should be noted that
both \cite{wessling1} and our calculations have the negative
parity states being more stable. The greater binding for the
quadratic (versus the constant) potential is natural since it is
significantly deeper.

We also analyze the case in which the short-distance potential is
simply set to zero.  For this case, the OPEP does not bind a
pentaquark for any channel.  In order for this potential to bind
without the aid of short-distance potential, $g_H$ would need to be
raised to unreasonably high levels, near 1 (approximately double the
extracted value), and in some cases larger than 2\@.  When realistic
mass differences between the vector and pseudoscalar mesons are
introduced, binding becomes weaker.  This mass splitting eliminates
binding for all channels with either type of potential we consider.

The heavy-meson coupling constant $g_H$ used in our analysis is
motivated by the results of a recent experiment by the CLEO
Collaboration that measured~\cite{cleo} the width of the $D^{*\pm}
\rightarrow D^{0}\pi^{\pm}$ decay.  The value of $g_H$ is
extracted from the width and found to be $\pm 0.59 \pm 0.07$. The
analogous decay process is energetically forbidden the in $B$
sector, preventing a direct extraction; therefore, we exploited
heavy quark symmetry and used the same value of $g_H$ for the $B$
sector.  Note, however, the uncertainty in the bottom sector due
to possible $1/m_Q$ corrections.  Accordingly, we also
investigated using a range of heavy-meson couplings and find the
same qualitative results.

These results depend upon the strength of the short-distance
potential.  Clearly, as these potentials become more strongly
attractive, the states are more likely to bind. As the potential
needed to bind deuterium may by anomalously small, a deeper constant
potential was also considered. Table~\ref{tab:Bmeson}, Col.\ C and
Table~\ref{tab:Dmeson}, Col.\ A show the results when the constant
potential is decreased from the depth needed to bind deuterium,
$-62.79$~MeV, to about 4 times as deep, $-276$~MeV. The deeper well
both produces more bound states and causes previously unbound states
to bind (In particular, the $D$ meson can form a bound state in the
deeper potential).

The choice of OPEP cutoff at $r \! = \! 1$~fm is arbitrary.  One does
not expect the OPEP to be valid for $r \! < \! 1$~fm, but the
effective cutoff might occur at somewhat larger $r$.
Table~\ref{tab:Bmeson}, Col.\ D and Table~\ref{tab:Dmeson}, Col.\ B
present the binding of states with a cutoff of 1.5~fm (the potential
depth is $-62.79$~MeV). The negative-parity states remain the only
bound ones, but the binding is now stronger, and the $D$ meson
binds. These fluctuations in strength of binding indicate the
importance of the short-distance physics to the heavy pentaquark
formation.

\section{Discussion} \label{concl}

Despite our general argument using the large $N_c$ and large $m_Q$
combined limit that the long-range OPEP is sufficient to bind
pentaquarks, we find in our class of models that, if a heavy
pentaquark binds at all due to one-pion exchange, it does so weakly in
a few channels and depends in a nontrivial way upon the details of the
short-range interaction.  The main implication is obvious: In the real
world, $1/N_c$ and $1/m_Q$ corrections can be substantial.  Indeed,
they are large enough to render unreliable even qualitative
predictions about heavy pentaquarks based upon the combined limit.

Given this somewhat unhappy result, the most important question is
whether or not heavy pentaquarks do in fact bind to form stable states
under strong interactions, and if so, whether only very weakly-bound
states occur, such as the ones seen here.  Both of these questions
remain open.  We simply do not know enough about the short-distance
part of the effective potential to provide a definitive answer.  An
optimistic view is that the short-distance interaction between the
heavy meson and the nucleon is likely to be more attractive than that
between nucleons, which has a strong repulsive core.  This argument is
particularly plausible if one views at least part of the repulsive
core between nucleons to arise due to the Pauli principle between
overlapping nucleon wave functions; this effect is greatly reduced in
the interaction between a nucleon and a heavy meson.  If it is true
that the short-range effective potential between the heavy meson and
the nucleon is significantly more attractive than the analogous
nucleon-nucleon case, then it is quite likely that heavy pentaquarks
form stable, tightly-bound states.

Finally, we address the question of why the qualitative
prediction of the combined large $N_c$ and large $m_Q$ limits is
insufficient.  At first sight this may seem surprising, since
both the $1/N_c$ and $1/m_Q$ expansions have proven to be
predictive in many situations. One must remember, however, that
the quality of a systematic expansion depends on coefficients as
well as the expansion parameter, and the size of these
coefficients depends on the observable being studied. If some
observable has ``unnaturally'' large coefficients, then the
expansion can easily fail unless the expansion parameter is
extremely small.  This view is echoed in \cite{kopeliovich}. The
relevant question is whether one ought to expect ``unnaturally''
large corrections to the leading behavior.

In retrospect, it is perhaps not so surprising that combined expansion
is insufficient here.  One can make an analogous argument, based
entirely upon $1/N_c$ counting, that both the deuteron and the $^1 \!
S_0$ two-nucleon channel ought to be deeply bound and have a large
number of bound states: Both the effective interaction between
nucleons and the masses of the two nucleons grow as $N_c^1$.  However,
as has been stressed elsewhere~\cite{cohen2}, this argument fails for
smaller values of $N_c$.  Similarly, numerous doubly-heavy
strongly-bound tetraquarks ought to exist in the heavy quark limit:
The effective interaction between heavy mesons is independent of the
heavy quark mass and scales as $1/(N_c m_Q)$.  However, as discussed
in Ref.~\cite{manohar} and based upon models similar to those studied
here, it is questionable whether they are bound for finite $m_Q$.
Evidently, the coefficients describing interactions between hadrons
can in some qualitative way be sufficient to weaken significantly
results one would naively expect directly from the $1/N_c$ or $1/m_Q$
expansions, yielding very large corrections to the leading-order
results for real-world parameters.  Why this is so is one of QCD's
more intriguing mysteries.

In conclusion, we showed that heavy pentaquarks must exist in combined
large $N_c$ and large $m_Q$ limit, and that they form multiplets of
SU(4)$\, \times \,$O(8)$\, \times \,$SU(2).  We constructed a one-pion
exchange potential between a nucleon and a heavy meson, and solved
coupled nonrelativistic Schr\"odinger equations, obtaining bound
states.  Some weakly-bound states do exist in some models, but their
existence depends upon unknown short-distance physics.  The lack of
binding emphasizes that the real world is too far from the idealized
world of large $N_c$ and large $m_Q$ to render the expansions robust
for these observables.  In order to deduce whether or not heavy
pentaquarks exist requires a more complete understanding of the
short-distance physics than is presently known.

{\it Acknowledgments.}  T.D.C.\ and P.M.H.\ were supported by the
D.O.E.\ through grant DE-FGO2-93ER-40762; R.F.L.\ was supported by the
N.S.F.\ through grant PHY-0140362.

\appendix

\section{Bound States in Quantum Mechanics} \label{QM}

Consider a smoothly varying potential $V(r)$ that vanishes as $r
\! \to \! \infty$.  If $V(r)$ is nonsingular and has an attractive
region, it must possess a minimum at some $r_0$.  In the
neighborhood of $r_0$ the potential is approximately harmonic
[{\it i.e.}, $V(r) \simeq \frac k 2 (r-r_0)^2$].  Therefore, if
the wave function is for some reason localized near the minimum,
then the system can be approximated as a harmonic oscillator. For
large reduced mass $\mu$ the kinetic energy operator is small,
and minimizing the wave function's curvature forces its
localization near $r \! = \! r_0$, as desired.  The harmonic
oscillator potential has an infinite number of bound states,
separated by multiples of $\omega \!  = \! \sqrt{k/\mu}$.  Thus
we see that multiple bound states must exist for sufficiently
large $\mu$.  If the potential is singular (but not more singular
than $1/r^2$, so that a ground state exists), the large size of
$\mu$ localizes the wave function deep in the potential near the
singularity, again allowing plenty of room for bound states.

\section{Tables of Results} \label{tables}
This appendix focuses on our numerical results.  Table~\ref{tab:const}
lists the parameters used in the calculation.
Table~\ref{tab:potential} summarizes the potentials that were used.
Table~\ref{tab:Bmeson} presents the energies of bound states for a $B$
meson binding with a nucleon, while Table~\ref{tab:Dmeson} presents
the same for a $D$ meson.

\begin{table}[ht]
\begin{tabular}{|c|c|}
\hline Quantity Name & Quantity Value\\
\hline
$g_A$ & 1.27\\
$f_\pi$ & 131 MeV\\
$g_H$ & $\pm$ 0.59\\
$m_\pi$ & 138 MeV\\
$m_N$ & 938.92 MeV\\
$m_B$ & 5279 MeV\\
$m_D$ & 1867 MeV\\
$\Delta_B$ & 46 MeV\\
$\Delta_D$ & 141 MeV\\
\hline
\end{tabular}
\caption{Constants used in bound-state calculations for heavy
pentaquarks.} \label{tab:const}
\end{table}

\begin{table}[ht]
\begin{eqnarray}
V_\pi (\vec{x})& = & \left\{ \begin{array}{lll}
(I^2-I_N^2-I_H^2)[S_{12}  V_T (r) +
       (K^2-S_N^2-S_H^2) V_c(r)] && r>r_0\\
V_1(r)\ \textrm{or}\ V_2(r) && r<r_0 \end{array} \right. \nonumber \\
V_c(r)& = & \frac{g_A g_H}{2\pi f_\pi^2} \frac{e^{-m_\pi r}}{3r} m_\pi^2 \nonumber \\
V_T (r)&=&\frac{g_A g_H}{2\pi f_\pi^2} \frac{e^{-m_\pi r}}{6r}
\left( \frac{3}{m_\pi^2 r^2} + \frac{3}{m_\pi r} + 1 \right) m_\pi^2 \nonumber\\
V_1(r) &=& V_0 \; (V_0 = -62.79\ \textrm{MeV or} -276\ \textrm{MeV})\nonumber\\
V_2(r) &=& -252.659\, \frac{\textrm{MeV}}{\textrm{fm}^2}\, r^2+541.321\, \frac{\textrm{MeV}}{\textrm{fm}} \, r -309.822\, \textrm{MeV}\nonumber\\
\end{eqnarray}

\caption{Potentials used in heavy pentaquark calculations.  The
labels are: total isospin $I$, nucleon isospin $I_N$, heavy meson
isospin $I_H$, tensor force $S_{12}$, tensor potential $V_T(r)$,
nucleon spin $S_N$, light quark in heavy meson spin $S_l$, sum of
nucleon spin and light quark spin $K$, central potential
$V_c(r)$. Numerical values are such that potentials are measured
in  MeV, distances in MeV$^{-1}$, unless noted otherwise. Both
$V_1(r)$ and $V_2(r)$ are central potentials. The parameters in
$V_2(r)$ were fixed by making the potential differentiable at $r_0$
and bind deuterium with the appropriate energy.}
\label{tab:potential}
\end{table}

\begin{table}[ht]
\begin{tabular}{|ccc|c|cc|cc|cc|cc|}
\hline
\multicolumn{3}{|c|}{Channel} & I & \multicolumn{2}{|c|}{A} & \multicolumn{2}{|c|}{B} & \multicolumn{2}{|c|}{C} & \multicolumn{2}{|c|}{D} \\
J & S & P & & + & $-$ & + & $-$ & + & $-$ & + & $-$\\
 \hline
 $\frac{1}{2}$ & $\frac{1}{2}$ & $-$ & 0 & 1.30 & 1.35 & 3.89 & 1.92,
3.62 & 139.38, 142.14 & --& 14.49, 16.01 & 15.46, 16.15\\
 && & 1 & -- & -- & 0.35 & 0.27 & -- & 139.38, 140.76& 15.32, 15.60 & 15.04, 15.46 \\ \hline
 $\frac{1}{2}$ & $\frac{1}{2}$ & + & 0 & -- & --& -- & --& 14.9, 32.39 & 4, 19.32, 46.5& -- & --\\
 && & 1 & -- & --& -- & --& 12.72, 18.22, 26.91 & 9.45& -- & --\\ \hline
 $\frac{1}{2}$ & $\frac{3}{2}$ & $-$ & 0 & 1.30 & 1.31& 3.89 & 3.67& 140.76 & 140.76& 15.87 & 15.32\\
 && & 1 & -- & --& -- & 0.26& 140.76 & 140.76& 15.04 & 15.32\\ \hline
 $\frac{1}{2}$ & $\frac{3}{2}$ & + & 0 & -- & --& -- & --& 32.15 & 3.35, 45.95& -- & --\\
 && & 1 & -- & --& -- & --& 12.12, 27.19 & 8.36, 22.08& -- & --\\ \hline
 $\frac{3}{2}$ & $\frac{1}{2}$ & $-$ & 0 & 1.42 & 1.31& 3.89 & 3.67& 140.76 & 140.76& 15.87 & 15.32\\
 && & 1 & -- & --& -- & 0.26& 140.76 & 140.76& 15.04 & 15.32\\ \hline
 $\frac{3}{2}$ & $\frac{1}{2}$ & + & 0 & -- & --& -- & --& 15.32, 18.49, 32.43 & 4.65& -- & --\\
 && & 1 & -- & --& -- &--& 12.80 & 17.25, 17.66, 22.91& -- & --\\ \hline
 $\frac{3}{2}$ & $\frac{3}{2}$ & $-$ & 0 & 1.42 & 1.25& 3.89 &3.67& 140.76 & 140.76& 15.87 & 15.32\\
 &&& 1 & -- & --& -- & 0.20& 140.76 & 140.76& 15.04 & 15.32\\ \hline
 $\frac{3}{2}$ & $\frac{3}{2}$ & + & 0 & -- & --& -- & --& 18.22, 32.29 & --& -- & --\\
 && & 1 & -- & --& -- & --& 4.18, 23.18 & --& -- & --\\ \hline

\end{tabular}

\caption{$B$ meson bound-state energies for each channel, where
$+$ and $-$ refer to relative sign of $g_A$ and $g_H$.  All energies
in MeV\@.  Column A: constant potential, $V_0=-62.79$~MeV and
$r_0=1$~fm; B: quadratic potential; C: constant potential,
$V_0=-276$~MeV and $r_0 = 1$~fm; D: constant potential,
$V_0=-62.79$~MeV and $r_0 = 1.5$~fm.}
\label{tab:Bmeson}

\end{table}

\begin{table}[hp]
\begin{tabular}{|ccc|c|cc|cc|}
\hline
\multicolumn{3}{|c|}{Channel} & I & \multicolumn{2}{|c|}{A} & \multicolumn{2}{|c|}{B}\\
J & S & P & & + & $-$& +& $-$\\
 \hline
 $\frac{1}{2}$ & $\frac{1}{2}$ & $-$ & 0 & 113.99, 110.4 & --& 7.36, 9.00 & 8.45, 9.27\\ &&& 1 & -- & 114.82, 115.78& 8.40, 8.79 & 8.16, 8.63\\ \hline
 $\frac{1}{2}$ & $\frac{1}{2}$ & + & 0 & 2.91 & 16& -- & --\\ &&& 1 & -- & --& -- & --\\ \hline
 $\frac{1}{2}$ & $\frac{3}{2}$ & $-$ & 0 & 117.3 & 116.2& 9.00 & 8.45\\&& & 1 & 115.23 & 115.23& 8.45 & 8.45\\ \hline
 $\frac{1}{2}$ & $\frac{3}{2}$ & + & 0 & 2.10 & 15.87& -- & --\\ &&& 1 & -- & --& -- & --\\ \hline
 $\frac{3}{2}$ & $\frac{1}{2}$ & $-$ & 0 & 117.3 & 116.20& 9.00 & 8.45\\ &&& 1 & 115.37 & 115.78 & 8.45 & 8.45\\ \hline
 $\frac{3}{2}$ & $\frac{1}{2}$ & + & 0 & 2.91 & --& -- & --\\&& & 1 & -- & --& -- & --\\ \hline
 $\frac{3}{2}$ & $\frac{3}{2}$ & $-$ & 0 & 117.3 & 116.20& 9.00 & 8.45\\ &&& 1 & 115.09 & 115.09& 8.45 & 8.45\\ \hline
 $\frac{3}{2}$ & $\frac{3}{2}$ & + & 0 & 2.53 & --& -- & --\\ &&& 1 & -- & --& -- & --\\ \hline
\end{tabular}

\caption{$D$ meson bound-state energies for each channel, where
$+$ and $-$ refer to the relative sign of $g_A$ and $g_H$.  All
energies in MeV\@.  Column A: constant potential, $V_0=-276$~MeV and
$r_0 = 1$~fm; B: constant potential, $V_0=-62.79$~MeV and $r_0 =
1.5$~fm.} \label{tab:Dmeson}

\end{table}

\end{document}